\documentclass[sigconf]{acmart}

\AtBeginDocument{%
  }

\copyrightyear{2025} 
\acmYear{2025} 
\setcopyright{cc}
\setcctype{by}
\acmConference[FAccT '25]{The 2025 ACM Conference on Fairness, Accountability, and Transparency}{June 23--26, 2025}{Athens, Greece}
\acmBooktitle{The 2025 ACM Conference on Fairness, Accountability, and Transparency (FAccT '25), June 23--26, 2025, Athens, Greece}\acmDOI{10.1145/3715275.3732080}
\acmISBN{979-8-4007-1482-5/2025/06}

\usepackage{colortbl}

\begin{document}

\title{Promising Topics for U.S.–China Dialogues on AI Risks and Governance}

\author{Saad Siddiqui}
\authornote{core contributors to this paper}

\affiliation{%
  \institution{Safe AI Forum}
  \country{United Kingdom}
}

\author{Lujain Ibrahim}
\authornotemark[1]

\affiliation{%
  \institution{University of Oxford}
    \country{United Kingdom}
}

\author{Kristy Loke}
\affiliation{%
  \institution{Independent}
  \country{United States}
}

\author{Stephen Clare}
\affiliation{%
  \institution{Centre for the Governance of AI}
  \country{United Kingdom}
}

\author{Marianne Lu}
\affiliation{%
  \institution{Stanford University}
  \country{United States}
}

\author{Aris Richardson}
\affiliation{%
  \institution{Centre for the Governance of AI}
  \country{United Kingdom}
}

\author{Conor McGlynn}
\affiliation{%
  \institution{Harvard University}
  \country{United States}
}

\author{Jeffrey Ding}
\affiliation{%
  \institution{George Washington University}
  \country{United States}
}

\renewcommand{\shortauthors}{Siddiqui et al.}

\begin{abstract}
Cooperation between the United States and China, the world's leading artificial intelligence (AI) powers, is crucial for effective global AI governance and responsible AI development. Although geopolitical tensions have emphasized areas of conflict, in this work, we identify potential common ground for productive dialogue by conducting a systematic analysis of more than 40 primary AI policy and corporate governance documents from both nations. Specifically, using an adapted version of the AI Governance and Regulatory Archive (AGORA) — a comprehensive repository of global AI governance documents — we analyze these materials in their original languages to identify areas of convergence in (1) sociotechnical risk perception and (2) governance approaches. We find strong and moderate overlap in several areas such as on concerns about algorithmic transparency,  system reliability, agreement on the importance of inclusive multi-stakeholder engagement, and AI's role in enhancing safety. These findings suggest that despite strategic competition, there exist concrete opportunities for bilateral U.S.-China cooperation in the development of responsible AI. Thus, we present recommendations for furthering diplomatic dialogues that can facilitate such cooperation. Our analysis contributes to understanding how different international governance frameworks might be harmonized to promote global responsible AI development.
\end{abstract}



\received{20 January 2025}
\received[accepted]{9 April 2025}

\maketitle

\section{Introduction}
Artificial intelligence (AI) systems developed in one country can profoundly shape outcomes around the world. As such, global cooperation is needed to effectively manage the transnational nature of AI development and deployment, especially in areas such as risk mitigation and safety standards \cite{pouget2024future}. Cooperation between the United States (U.S.) and China is particularly important as the two countries lead in global AI investment, research, development, and deployment, together accounting for the majority of global AI advancement \cite{horowitz2018artificial}.

Leaders in both the U.S. and China have already acknowledged the need for effective governance to address AI risks. The Biden administration’s 2023 ``Executive Order on the Safe, Secure, and Trustworthy Development and Use of Artificial Intelligence" (Biden AI EO) emphasizes that to manage substantial risks from AI, ``policies, institutions, and, as appropriate, other mechanisms" are required to test and evaluate systems before they are deployed \cite{biden2023executive}. Similarly, a resolution adopted at the third plenary session of the 20th Central Committee of the Communist Party of China states that the intention of the Chinese Communist Party (CCP) is to ``improve the mechanisms for developing and managing generative artificial intelligence" and ``institute oversight systems to ensure the safety of artificial intelligence" \cite{Welch_2024}. 

The two countries have also recognized the importance of international cooperation on AI governance and safety, with both signing the Bletchley Declaration at the 2023 AI Safety Summit \cite{oxford2023uk}. China has also co-sponsored the U.S.-led 2024 United Nations General Assembly resolution to promote safe, secure, and trustworthy AI systems for sustainable development \cite{Geneva_Graduate_institute_2024}. These official initiatives have been complemented by Track II diplomacy efforts, such as the International Dialogues on AI Safety, a series of discussions between scientists from both nations \footnote{List of dialogues here: https://idais.ai/dialogues/}.

Despite these developments, significant obstacles remain to U.S.-China cooperation on AI issues. Amid current geopolitical tensions and strategic competition between the two nations, leadership in advanced AI system development is considered critical for both national security and economic competitiveness \cite{wu2020technology}. Accordingly, the U.S. and China continue to make substantial investments in domestic AI projects, while seeking to control critical inputs for AI development. Most notably, the U.S. has implemented a series of export controls to restrict Chinese firms' access to advanced semiconductors crucial for AI development \cite{brockmann2022applying}.

However, in the past, states have been able to cooperate on issues of mutual interest even in the presence of fundamental disagreements. For example, during the Cold War, the U.S. and the Soviet Union cooperated on nuclear safety and security issues to prevent unintended nuclear detonations, showing that cooperation between rival powers is possible even on issues tied to strategic competition \cite{ding2024keep}. Such cooperation typically begins with sustained dialogue to identify shared interests and build mutual understanding. Building on this historical evidence that dialogue and cooperation can coexist with strategic rivalry, this paper identifies several key areas where U.S.-China dialogue on AI governance and safety could prove productive. Specifically, we conduct a direct analysis of primary documents in their original languages by researchers fluent in Chinese and English. Our work contributes to the FAccT community's ongoing examination of AI governance frameworks by providing a systematic analysis comparing how the world's two leading AI powers approach key sociotechnical challenges. While much attention has focused on points of conflict between U.S. and Chinese AI development, our approach focuses on revealing specific areas where shared interests and compatible governance frameworks could enable meaningful dialogue on responsible AI development.

The paper is structured as follows: we begin with a brief background on the broader U.S.-China AI relationship. This is followed by an outline of our methodology for selecting and analyzing key AI documents from both nations. We then present the results of our analysis of over 40 primary AI policy and corporate governance documents, examining various AI risks (e.g., limited user transparency, poor reliability) and governance approaches (e.g., setting up new institutions). Based on this analysis, we surface areas of overlap and present recommendations for promising areas of bilateral dialogue and potential cooperation. We conclude with a discussion of the limitations of our approach and opportunities for future research.

\textit{\textbf{Note:}  This research was conducted and written prior to the rescission of the Biden AI EO, which was/is critical piece in the U.S. AI policy landscape and is a significant part of our discussion here. We acknowledge that political landscapes are fragile and ever-shifting; we discuss these implications further in Section~\ref{sec:eo}.}

\section{Background}
We provide relevant background information on the U.S.-China AI relationship, primarily through (1) a review of existing comparative analyses of U.S. and Chinese AI policy, and (2) a brief outline of recent international cooperation on AI safety and governance which involved representatives from both nations. 

\subsection{Comparative analysis of U.S. and Chinese AI regulation}
Recent comparative analyses of U.S. and Chinese AI governance reveal a complex landscape of convergence and divergence. The philosophical and quantitative analyses by Hine and Floridi (2022) and Hine (2023) established that there exist shared aspirations, for example for equitable AI benefit distribution, though tensions between U.S. democratic values and global pluralistic governance remain significant challenges\cite{hine2022artificial, hine2023governing}. Building on this foundation, Chun et al. (2024) identify fundamental differences in regulatory philosophies, contrasting the U.S.'s market-driven approach with China's more centralized framework \cite{chun2024comparative}, while Giardini and Fritz (2024) find meaningful alignment on certain technical issues like algorithmic discrimination and cybersecurity requirements \cite{giardini2024anatomy}. However, Giardini and Fritz (2024) also note important distinctions in implementation, such as the U.S.'s prohibition-based approach versus China's licensing regime for certain AI systems, as well as divergent content moderation standards reflecting different sociopolitical priorities. 

Our work builds on and extends these analyses, particularly the methodological approach of Zeng et al. (2024) in categorizing government and corporate documents \cite{zeng2024ai}. However, where their focus was on developing a common risk taxonomy, our analysis specifically aims to identify practical opportunities for bilateral cooperation despite acknowledged philosophical differences.

\subsection{Continuities and discontinuities in U.S. AI policy} \label{sec:eo}
Our analysis was completed before the rescission of the Biden AI EO, so we must acknowledge its implications. Despite changes between the first Trump administration, the Biden administration, and the current second Trump administration in their attitudes towards AI safety and governance, both in terms of domestic priorities and engagement with rivals like China, there is also a significant degree of continuity. The largest areas of continuity are the emphasis placed on U.S. engagement in global standards-setting, restricting China’s access to advanced AI chips through export controls, and a limited but important willingness to engage with the Chinese government in critical intergovernmental negotiations \cite{eo2023executive, whitehouse2025}.  

Biden's AI EO emphasized responsible AI principles while Trump's focused on competitiveness and deregulation, yet maintained key initiatives like the U.S. AI Safety Institute. In standards-setting strategy, Trump's first administration positioned NIST to compete with China through international bodies, while Biden created exemptions allowing American companies to engage with Chinese entities for standards development \cite{nist2019}. Semiconductor export controls began under Trump in 2020 and were expanded four times by Biden's Commerce Department, with Trump's second administration signaling simplification rather than elimination of these restrictions \cite{feldgoise2024}. Both administrations pursued high-level diplomatic engagement, with Biden-Xi talks yielding agreements on nuclear decision-making control, and Trump's administration continuing dialogue through Treasury channels \cite{renshaw2024, revill2025}. The most notable shift thus far has been in framing, with the Trump administration avoiding ``safety" terminology, exemplified by JD Vance's declaration that ``The AI future will not be won by hand-wringing about safety. It will be won by building" \cite{bristow2025}. 

As the second Trump administration continues to implement its AI agenda, policy will undoubtedly evolve in ways that are difficult to predict, though the fundamental U.S. interests in maintaining technological leadership, engaging in standards bodies, restricting strategic competitors' access to advanced capabilities, and pursuing diplomatic channels are likely to remain consistent drivers of policy.

\subsection{Evidence from diplomatic processes}
Recent multilateral engagements reveal emerging patterns of U.S.-China cooperation on AI governance. The 2023 Bletchley Declaration, signed by both nations, acknowledged shared concerns about AI's risks to human rights, privacy, fairness, and potential catastrophic harm from misuse or loss of control \cite{bletchley_declaration_2023}. While China's non-participation in the 2024 Seoul Summit's intergovernmental statement suggests some limitations to this consensus, major Chinese AI developer Zhipu AI's commitment to the Seoul corporate pledges indicates continued private sector engagement \cite{birch2024key}. Both nations also supported the 2024 United Nations (UN) General Assembly resolution on AI, which highlighted shared interests in leveraging AI for sustainable development while acknowledging concerns about digital divides and structural inequalities. Bilateral progress has also emerged, with both countries holding intergovernmental AI talks in Geneva and reaching an agreement on preventing AI control over nuclear weapons decisions \cite{zhang2024us, renshaw2024biden}.

Track II diplomacy - unofficial dialogue between non-state actors from different countries to build relationships and solve problems - has further advanced these cooperative efforts. Henry Kissinger's 2023 discussions with Chinese President Xi Jinping emphasized the need for high-level international cooperation on AI risks \cite{davis2024back}. Concurrently, the International Dialogues on AI Safety (IDAIS) produced two significant consensus statements: the Ditchley Statement calling for coordinated global action on AI risks, and a subsequent Beijing Statement establishing specific technological ``red lines" including autonomous replication and deception of regulators \footnote{International Dialogues on AI Safety, https://idais.ai/}. Given the breadth and severity of AI risks acknowledged in these various forums, and the demonstrated potential for bilateral engagement, there is a clear need to further identify specific areas for dialogue where meaningful U.S.-China cooperation could be possible. Our analysis attempts to systematically address this need by focusing on areas of shared risk perceptions and shared governance approaches. 

\section{Analysis methodology}
Our methodology combines systematic document analysis with qualitative coding to identify areas of potential U.S.-China cooperation on AI governance. We analyzed 44 primary documents from both nations, including government policies, corporate whitepapers, and model release papers, coding them for specific (1) AI risks and (2) governance approaches. Then, we use a standardized grading framework to evaluate the documents to assess the degree of overlap between U.S. and Chinese perspectives. The documents were collected in April 2024, and all the analyses were completed in April, May, and June of 2024. 

\subsection{Identification and selection of relevant documents}
Our dataset comprises primary documents from both government and corporate entities in the U.S. and China. Government documents were systematically collected from two authoritative sources: the Digital Policy Alert database \footnote{https://digitalpolicyalert.org/activity-tracker} and Concordia AI's comprehensive 2023 State of AI Safety in China report \footnote{https://concordia-ai.com/wp-content/uploads/2023/10/State-of-AI-Safety-in-China.pdf}. These sources provided regulatory documents from key agencies and institutions within both nations' governmental frameworks. To complement official policy documents, we conducted a structured review of corporate governance materials, including technical whitepapers and model release papers from major AI companies in both nations, focusing on their articulated approaches to AI governance and safety.

\subsubsection{Inclusion and exclusion criteria}
We established specific temporal and jurisdictional boundaries for our corpus. To capture contemporary perspectives on generative AI governance, we primarily focused on documents released in 2023 or later. However, we extended this temporal boundary to 2021 in cases of limited data availability, particularly for Chinese corporate policy documents. We also made strategic exceptions to include seminal pre-2023 documents that remain foundational to current AI governance frameworks, such as China's recommendation algorithm regulations.

For government sources, we restricted our analysis to official, national-level documents to ensure comparability and relevance to international cooperation potential. This meant excluding sub-national policies (e.g., from Beijing or California) despite their domestic influence. We made a strategic exception for two widely-discussed Chinese AI law proposals due to their potential significance for national policy direction.

While corporate documents do not fully represent national policy position, we include them in our analysis as there are several avenues through which key corporations influence AI policy: by participating in national policy discussions and standards-setting bodies, providing testimony at hearings, creating policy connections through revolving doors, and publishing governance frameworks that anticipate or respond to national AI policies. Moreover, corporate sources, particularly some of the whitepapers in our dataset, provide a unique window into how regulations are being applied in practice by companies. We employed targeted sampling of model release papers from leading AI companies in both nations, selecting documents that provided detailed insights into safety and governance approaches. We supplemented this with select Chinese corporate whitepapers to better understand the corporate interpretation and implementation of national regulations.

To maintain focus on authentic domestic policy positions, we deliberately excluded international declarations, think-tank reports, and judicial decisions. This choice reflects our emphasis on analyzing how each nation articulates its AI governance priorities to domestic audiences rather than international stakeholders.

\subsection{Coding framework and process}
Our analysis employed a modified version of the Georgetown Center for Security and Emerging Technology's AGORA taxonomy, a comprehensive framework for categorizing AI risks and governance approaches\footnote{https://eto.tech/tool-docs/agora/}. While AGORA was originally designed for document-level analysis, we adapted it for more granular span-level coding, augmenting the taxonomy with additional categories and refined definitions to capture nuanced policy positions. The framework encompassed multiple dimensions of AI risk (such as bias and AI-generated content risks) and governance approaches (such as institutional mechanisms), with each dimension having standardized definitions and exemplars.

\subsubsection{Implementation and quality control}
Researchers conducted line-by-line analysis of each document, applying the adapted taxonomy to specific text spans. Our coding schema allowed for multi-dimensional tagging, enabling individual text segments to be coded for both risk and governance dimensions, with multiple entries created when spans addressed multiple categories. To ensure coding consistency, we implemented a multi-stage review process: researchers explicitly flagged uncertain cases for team review, and an editor conducted systematic sample checks across the entire corpus to maintain methodological consistency.

While we prioritized analytical consistency through standardized category definitions and illustrative examples, we acknowledge the inherently interpretive nature of qualitative coding. Rather than employing formal inter-coder reliability metrics, our approach relied on researcher expertise and collaborative review processes to ensure analytical rigor. Two authors independently categorized each case of overlap, and disagreements were resolved through discussion until consensus was reached. 

\subsection{Dataset description}
Tables~\ref{tab:tab1}, ~\ref{tab:tab2}, and ~\ref{tab:tab3} provide a breakdown of the dataset entries by type of issuing entity, showing the number of documents, total number of quotes, and average number of quotes per document. 
\begin{table*}
\caption{Breakdown of the documents in the dataset by type of issuing entity}
\label{tab:tab1}
\begin{tabular}{lrrrrrr}
\toprule
\textbf{Country of issuance} & \textbf{Corporate} & \textbf{Executive (Party)} & \textbf{Executive (State)} & \textbf{Legislature} & \textbf{Others} & \textbf{Grand Total} \\
\midrule
China & 13 & 3 & 11 & & 3 & \textbf{30} \\
U.S. & 3 & & 9 & 1 & 1 & \textbf{14} \\
\textbf{Grand Total} & \textbf{16} & \textbf{3} & \textbf{20} & \textbf{1} & \textbf{4} & \textbf{44} \\
\bottomrule
\end{tabular}
\end{table*}

\begin{table*}
\caption{Breakdown of the quotes in the dataset by type of issuing entity}
\label{tab:tab2}
\begin{tabular}{lrrrrrr}
\toprule
\textbf{Country of issuance} & \textbf{Corporate} & \textbf{Executive (Party)} & \textbf{Executive (State)} & \textbf{Legislature} & \textbf{Others} & \textbf{Grand Total} \\
\midrule
China & 115 & 25 & 57 & & 44 & \textbf{241} \\
U.S. & 22 & & 161 & 2 & 4 & \textbf{189} \\
\textbf{Grand Total} & \textbf{137} & \textbf{25} & \textbf{218} & \textbf{2} & \textbf{48} & \textbf{430} \\
\bottomrule
\end{tabular}
\end{table*}

\begin{table*}
\caption{Breakdown of the average number of quotes per document in the dataset by type of issuing entity}
\label{tab:tab3}
\begin{tabular}{lrrrrrr}
\toprule
\textbf{Country of issuance} & \textbf{Corporate} & \textbf{Executive (Party)} & \textbf{Executive (State)} & \textbf{Legislature} & \textbf{Others} & \textbf{Grand Total} \\
\midrule
China & 9 & 8 & 5 & & 15 & \textbf{8} \\
U.S. & 7 & & 18 & 2 & 4 & \textbf{14} \\
\textbf{Grand Total} & \textbf{9} & \textbf{8} & \textbf{11} & \textbf{2} & \textbf{12} & \textbf{10} \\
\bottomrule
\end{tabular}
\end{table*}

\subsection{Corpus composition and distribution}
The corpus exhibits an asymmetry in document distribution, with Chinese sources comprising approximately twice the number of U.S. documents. This disparity reflects two key factors in our sampling strategy: first, the inclusion of a broader range of corporate documents from China's more distributed AI industry landscape, compared to our focused sampling of model cards from the three established U.S. frontier AI companies (OpenAI, Anthropic, and Google DeepMind); second, the incorporation of additional Chinese state and party documents to capture the evolutionary nature of Chinese AI governance, particularly its connection to pre-2023 regulatory frameworks such as the algorithm registry.

Despite the numerical disparity in source documents, the extracted quote count remains relatively balanced between nations, with notable variations across document categories. Quote density analysis reveals significant variation across document categories, primarily driven by document length and content relevance. Chinese corporate documents yielded a higher quote density due to the comprehensive nature of corporate whitepapers, while U.S. executive branch documents, particularly the Biden AI Executive Order, similarity provided a high quote density of relevant U.S. policy positions. These source concentrations also reflect the distinct institutional structures through which each nation's actors articulate their AI governance approaches.

\subsection{Analysis}
Following the coding process, we conducted a systematic cross-national comparison of AI risk perceptions and governance approaches using a tripartite evaluation framework (summarized in Table~\ref{ref:tab1}). We developed specific criteria to categorize the degree of convergence between U.S. and Chinese positions as strong (indicating consistent cross-national concurrence), moderate (reflecting partial alignment with notable divergences), or weak (suggesting fundamental differences in understanding or emphasis). This structured evaluation was applied systematically across each risk category and governance strategy in our taxonomy. Our findings present a detailed analysis of these convergence patterns demonstrating how specific policy positions and governance approaches align or diverge across the two contexts.

\definecolor{darkgreen}{rgb}{0.3,0.8,0.3}  \definecolor{lightgreen}{rgb}{0.9,1,0.9}       
\definecolor{lightyellow}{rgb}{1,1,0.9}        

\begin{table}
\caption{Criteria for assessing degrees of overlap between U.S. and Chinese documents}
\begin{tabular}{lp{0.6\columnwidth}}
\toprule
\textbf{Degree of overlap} & \textbf{Criteria} \\
\midrule
\rowcolor{darkgreen}\color{white}
Strong & Most, if not all, of the material found in the dataset shows concurrence between U.S. and Chinese understanding of a given risk or governance approach \\
\midrule
\rowcolor{lightgreen}
Moderate & While there is some overlap in understanding, there are also key differences that we could identify in the way concepts were raised \\
\midrule
\rowcolor{lightyellow}
Weak & Majority of the evidence pointed towards differing understandings of emphases on a given issue \\
\bottomrule
\end{tabular}
\label{ref:tab1}
\end{table}
In the next section, we share findings from our analysis of the 44 primary documents in our dataset. We specifically analyse how these documents discuss (1) AI risk perceptions (defined in Appendix Table~\ref{tab:risks}) and (2) AI governance approaches (defined in Appendix Table~\ref{tab:governance}). A list of acronyms of government bodies and documents can also be found in the Appendix. We restrict our analysis to areas of high and moderate overlap, as they present the most promising directions for future bilateral dialogues. The findings are summarized in Table~\ref{tab:summary}.

\begin{table}
\caption{Summary of Overlap in Risk Perceptions and Governance Approaches}
\label{tab:summary}
\begin{tabular}{llp{0.5\columnwidth}}
\toprule
\textbf{Degree} & \textbf{Domain} & \textbf{Areas of Overlap} \\
\midrule
\rowcolor{darkgreen}\color{white}
Strong & Risks & Limited user transparency; Poor reliability \\
\rowcolor{darkgreen}\color{white}
& Governance & Pro-safety role of AI systems; Stakeholder convening \\
\midrule
\rowcolor{lightgreen}
Moderate & Risks & Lack of robustness; Bias and discrimination; Lack of interpretability and explainability; Dangerous capabilities; Weak cybersecurity \\
\rowcolor{lightgreen}
& Governance & External auditing; Licensing and registration; Watermarking; Model evaluations; Adversarial testing \\
\midrule
\rowcolor{lightyellow}
Weak & Risks & Privacy \\
\rowcolor{lightyellow}
& Governance & Risk assessments; Information disclosure; Input controls; Liability; Pilots and testbeds \\
\bottomrule
\end{tabular}
\end{table}

\section{Risks}
With respect to risks, our analysis suggests that there is at least moderate overlap on \textit{all major risk categories} we tracked, except for privacy, which we classify as having ‘weak’ overlap. 

\subsection{Strong overlap} \subsubsection{Limited user transparency:} The NIST Risk Management Framework from the U.S. describes transparency as foundational for accountability, using it to refer to the extent to which information about an AI system and its outputs is available to a user \cite{nist2023artificial}. Chinese documents largely have a similar understanding. An AI safety standard (TC260-003) calls for more transparency, with specific requirements for disclosure of certain contextual information about models \cite{sac2024basic}. Older Chinese recommendation algorithm regulations from 2021 refer to transparency in a slightly different context, with requirements for service providers to provide users with information about rules used for content recommendation, similar to the NIST definition above \cite{cac2021provisions}.

\subsubsection{Poor reliability}: NIST via its Risk Management Framework in the U.S., frames reliability as ‘the ability of an item to perform as required, without failure,’ calling reliability a goal for overall correctness of an AI system \cite{nist2023artificial}. The Ministry of Industry and Information in China and a related think-tank (the CAICT) both mention reliability in a similar way, and further elaborate on it as an area to be standardized, with specific reliability testing required under a machine learning cybersecurity standard \cite{miit2024guidelines, caict2024registration}. 

\subsection{Moderate overlap}
\subsubsection{Lack of robustness:} in the U.S., NIST’s Risk Management Framework refers to robustness as ‘the ability of a system to maintain its level of performance under a variety of circumstances.’ Some understanding of robustness is apparent in Chinese sources, though it is unclear if the exact same definition is in use \cite{nist2023artificial}. Alibaba’s 2023 AI Governance whitepaper and a 2021 industry whitepaper both refer to robustness as part of new issues brought about by the black-box nature of generative AI models \cite{alibaba2023white, nqsic2023white}.

\subsubsection{Bias and discrimination:} Bias and discrimination is referenced by both Western and Chinese sources in some similar ways, but with notable differences. The White House \& FTC in the U.S., as well as the Chinese standards-setting body TC260 and two major AI companies Baichuan and Alibaba have all variously expressed concerns around discrimination and disinformation leading to societal harm \cite{alibaba2023white, baichuan2023baichuan, sac2024standards, ftc2024ai, eo2023executive}. These documents share a similar scope, both discussing discrimination/bias along some similar lines, notably race/ethnicity, sex, and religion/beliefs \cite{eeoc2023select,cac2023interim}. U.S. documents, particularly many parts of the Biden AI EO, discuss bias in relation to model use in decision-making in practice - including the responsibility of the decision-maker and the vendor to ensure adequate measures are in place to mitigate bias \cite{eo2023executive}. Chinese companies similarly point to the risk that bias introduces in decision-making processes. However, bias and discrimination are contextual and therefore understood in different ways by different actors. For example, while AI labs from both China and the U.S. test their models against bias benchmarks, labs like Baichuan acknowledge that their models may not fully account for biases relevant in non-Chinese contexts \cite{baichuan2023baichuan,google2023gemini,anthropic2024claude}. 

\subsubsection{Lack of interpretability and explainability:} U.S. and Chinese documents both acknowledge the importance of ensuring interpretability and explainability (I\&E). One key difference is that the U.S. appears more optimistic about interpretability and explainability–related governance mechanisms, while Chinese documents focus on I\&E as an inherent problem for AI systems. However, this does not mean that Chinese entities view I\&E as unsolvable. In the U.S., documents from NIST mention I\&E \cite{nist2023artificial}. Meanwhile, several Chinese documents, from various actors (Alibaba, CAC, SAC, MIIT) bring up I\&E \cite{alibaba2023white, cac2021provisions, sac2024standards}. Chinese actors tend to use the terms to describe the problems of the black-box nature of algorithms and inherent challenges for AI safety, with AI company Sensetime even referring explicitly to an “interpretability risk” resulting from our limited ability to understand algorithms \cite{sensetime2022ai}. That said, MIIT does relate I\&E to new governance mechanisms. 

\subsubsection{National-security relevant capabilities (also known as `Dangerous capabilities'):} U.S. labs clearly state the need for `dangerous capability' evaluations in their technical reports, model cards, and responsible scaling policies \cite{openai2023gpt}. Chinese labs have shown awareness of these developments in whitepapers \cite{tencent2024large}, with Chinese state-linked think-tanks running model evaluations beginning to include some types of dangerous capability evaluations as part of their benchmarks \cite{ding2024chinai}. 

With regard to dangerous capabilities linked to chemical, biological, radiological or nuclear (CBRN) weapons development, there is quite strong overlap. The U.S. frequently mentions CBRN risks, focusing on biological and chemical domains \cite{eo2023executive}. Chinese actors' mention of bio-chemical risks also occur in the documents reviewed albeit not to the same degree as in the U.S. documents. An updated AI safety standard by the TC260 cybersecurity standardization committee included mention of the use of AI to create chemical or biological weapons in 2024 (though this was removed in a later draft) \cite{sac2024basic}. Beyond that, hazardous chemicals are one of over 20 items that CAICT is evaluating models for. It does appear that Chinese actors tend to view these capabilities under the well-established lens of ‘dangerous content,’ as the CAICT report puts hazardous chemicals under the category content security and subcategory violation of laws and regulations alongside issues like gambling and pornography. In contrast, the Biden AI EO frames chemical and biological risks more broadly, linked to the differential risk added by AI systems in aiding actors in developing CBRN weapons, relative to the internet and weighed against the defensive advantages granted by the same AI systems \cite{eo2023executive}. 

With respect to cyber capabilities of AI systems, there is also some common ground. The Biden AI EO calls attention to cybersecurity as an area that AI could cause harm to, including through the development of autonomous cyber capabilities \cite{eo2023executive}. In China, the TC260 Basic Safety Requirements for Generative Artificial Intelligence Services refers to the possibility that AI may be used to write malware \cite{sac2024basic}. However, the U.S. specifically expresses greater concern about inadvertent advancement of adversary cyber capabilities. In an Executive Order prohibiting investments in certain national security technologies and products in specific countries of concern, it expresses concerns about transactions that advance the cyber-enabled capabilities of ‘countries of concern’, including AI products\cite{EO14105}.

With regards to persuasion as a dangerous capability, there appears to be limited recognition in China, while there is much more widespread concern in the U.S.. In the U.S., concern is explicitly flagged in the OpenAI Preparedness Framework and other technical documents from AI labs (although not in the Biden AI EO) \cite{google2023gemini,openai2023gpt}. Despite the dissimilarity, Chinese AI regulations have always been concerned with algorithms that have public mobilization or social opinion properties, suggesting that mass persuasion via AI is or could become a concern of the Chinese state as well \cite{cac2023interim}. The U.S. does appear to be concerned about the use of federal data to enhance dangerous capabilities, while an equivalent concern was not found in our Chinese documents. The Biden AI EO expresses specific concern that federal data may aid in the development of CBRN or autonomous cyber capabilities, and directs the Chief Data Officer Council to draw up guidelines on performing security reviews. The EO also directs the Department of Energy to develop tools to understand and mitigate security risks from AI \cite{eo2023executive}.

With regards to other dangerous capabilities, we find some evidence of Chinese concern around model autonomy, self-replication, and evasion of human control or oversight but such capabilities are undeniably a much more widely discussed issue in the U.S. \cite{eo2023executive}. 

\subsubsection{Weak cybersecurity:} Both countries express concerns about weakened cybersecurity, which can be roughly categorised into concerns related to theft of model weights and general concerns about software vulnerabilities. 

There is a common ground on the importance of preventing theft or undesired access to model weights. The Biden AI EO and White House voluntary commitments, in combination, oblige developers to disclose the ownership and possession of any dual-use foundation model weights, alongside the physical and cybersecurity measures taken to safeguard these from theft \cite{eo2023executive, whitehouse2023fact}. As early as 2021, an industry report from several Chinese AI labs noted the possibility of model theft \cite{nqsic2021white}, with a more recent Tencent Large Model Safety and Security report describing the need to prevent model information leakage within an organization, alongside a broader suite of measures to protect model weights \cite{tencent2024large}. A Chinese standards-setting body also calls on AI service providers to separate inference and training environments to prevent improper access and information leakage \cite{sac2024basic}.

There is less clear common ground on general cybersecurity of systems related to AI development and deployment. The only U.S. documents to mention this topic were the Biden AI EO, which calls on developers to share software vulnerabilities that have been discovered and any known exploits associated with the vulnerabilities, and the Cybersecurity and Infrastructure Security Agency (CISA) notice on the obligation of AI systems to be secure by design, which calls for AI products to be built in a way that reasonably protects against malicious cyberattacks \cite{eo2023executive, lai2023software}. In China, this issue generally receives more widespread attention. A Tencent Large Model Security and Safety Report calls attention to vulnerabilities in open-source platforms used in parts of the machine learning pipeline, suggesting that that risk component library be constructed to monitor and flag any security risks introduced during training \cite{tencent2024large}. Government standards documents also flag cybersecurity risks in infrastructure. An AI computing information security platform framework issued by the TC260 standard-setting body argues that platforms must design their own security functions such that the platform itself provides a secure computing environment and does not become a weak link in cyber attacks \cite{sac2023artificial}. 

\section{Governance approaches}
With respect to governance approaches, we find that there is less overlap, with the majority of governance approaches showing moderate or weak overlap.

\subsection{Strong overlap}
\subsubsection{Use of AI for pro-safety purposes:} There is significant overlap in understanding that AI systems can be helpful for safety in a variety of ways. In the U.S., the Biden AI EO calls on the Secretary of Defense and Secretary of Homeland Security to identify and pilot ways to use AI to discover and fix critical vulnerabilities in software systems and networks \cite{eo2023executive}. In calling on the Department of Homeland Security (DHS) to evaluate CBRN threats from AI, the Biden AI EO also asks them to consider the benefits of application of AI to counter CBRN threats.  On the other hand, in China, a recent whitepaper by Tencent in 2023 identifies that AI can be used to identify and predict cyber threats (e.g., by monitoring web traffic and flagging anomalous patterns) \cite{tencent2024large}. Chinese AI company Zhipu, in a technical release paper for their GLM-130B model, points to promoting LLM-inclusivity as a way to defend against potential harms, instead of restricting access to LLMs \cite{zeng2023glm}. The CAICT Safety Model Evaluation standard released in 2024 also explicitly points to their use of AI models in conducting evaluations \cite{caict2024authoritative}.

\subsubsection{Convening:} There is strong overlap between the U.S. and China when it comes to the convening of different stakeholders to share feedback or participate in AI's development or deployment. In the U.S., directives for convening come largely from the Biden AI EO, which focuses on the following types of convenings in particular: interagency councils, public participation and community engagement, external stakeholder engagement (e.g., private sector, academia, civil society), and consulting with experts, labs, and AI evaluators \cite{eo2023executive}. In China, regulations such as the Deep Synthesis regulations, call for industry-led self-governance. This convening and collaboration is further evidenced by the list of authors for various Chinese standards, spanning academia, industry, and government do suggest that multi-stakeholder convening does take place \cite{cac2022provisions}. Calls for the convening of AI stakeholders appear in draft laws written by experts and legal scholars, rather than codified regulations. For example, the draft law from the Chinese University of Political Science and Law stipulates the creation of an expert committee (including experts in technology, law, and ethics) that provides support for AI safety work \cite{cls_ai_law}. The CASS draft model law also focuses on an AI government mechanism involving stakeholders from the government, public, and corporate sector \cite{cass2024model}. 

\subsection{Moderate overlap}
\subsubsection{Governance development:} In general, it is apparent that both nations are invested in developing governance systems for AI. Here we focus on approaches to international governance development, domestic governance development, and the creation of new institutions. Both the U.S. and Chinese governments have stated that they want to collaborate with international partners in governing AI, but given the broad scope of such statements, it seems unclear if there is any actual overlap here. For instance, the Biden AI EO states that the U.S. will seek partnerships with other countries on building safeguards \cite{eo2023executive}, while the Cyberspace Administration of China (CAC) has said that China will “carry out international exchanges and cooperation in an equal and mutually beneficial way” with other nations \cite{cac2023interim}. 

Both the U.S. and China also call for the development of new domestic governance approaches. The Biden AI EO calls upon parts of the government to develop new frameworks and institutions \cite{eo2023executive}, while Chinese documents, such as the two model laws, also call for the establishment of various governance mechanisms and in the case of the Chinese Academy of Social Science (CASS) draft model law, call for the establishment of a new centralized administration for managing AI governance \cite{cass2024model}.  Differences here include that some of the Chinese documents suggest that there is more obvious pressure put on companies to be responsible to their users. This focus on the responsibility of companies to users is largely absent from the U.S. texts on governance development. 

The U.S. currently appears to be more focused on creating new institutions, mostly within the executive branch. Besides the U.S. AI Safety Institute within the Department of Commerce, this also includes the establishment of the four new National AI Research Institutes, a coordinating office within the Department of Energy, a task force at the Department of Health and Human Services (DHHS), and a Research Coordination Network focused on advancing privacy research \cite{eo2023executive}. 

In China, only one of the two AI law proposals written by legal experts from the Chinese Academy of Social Sciences and industry experts has called for the creation of the China Administration of AI, which would be an overall central authority, sitting under a leading small group on AI (a party institution) \cite{cass2024model}. The other AI law proposal in China does not call for the creation of a new institution, but rather calls for working through existing institutions. More recently, however, the Third Plenum decision in China has called for instituting new oversight mechanisms related to safety, but it is unclear if this will be a new institution or part of existing institutions. 

\subsubsection{Technical solutions:} Documents on both the U.S. and China focus on the use of technical solutions as a strategy to govern particular AI risks. There exists significant overlap between the two countries around the need for content provenance and labeling to ensure information traceability. The Biden AI EO emphasizes capturing AI models and their dependencies in "software bills of materials" \cite{eo2023executive}, and U.S. corporate documents have pledged to develop labeling mechanisms for AI-generated content \cite{whitehouse2023fact}. Similarly, Chinese draft laws, amongst other standards and laws, mention the need for AI providers to add implicit identifiers and develop traceability mechanisms for users \cite{cls_ai_law}. On the U.S. side, there is a greater focus on privacy-enhancing technologies (PETs) than on the Chinese side. The White House EO refers to PETs being critical to protecting users’ privacy and combating broader legal and societal risks that result from the improper collection and use of people’s data \cite{eo2023executive}. In Chinese documents we found less extensive mention of PETs, though a Alibaba Whitepaper did argue that technical mitigations should be actively adopted to reduce the collection of personal information \cite{alibaba2023white}. 

\subsubsection{Evaluations:} There is some basic agreement that pre-deployment and post-deployment evaluations and monitoring are necessary, but the focus of evaluations appears to be different in the U.S. and Chinese documents. The Biden AI EO refers to the need for the government to provide guidance on benchmarks and evaluations, both with respect to pre-deployment testing and post-deployment performance monitoring to ensure that systems function as intended and are resilient to misuse \cite{eo2023executive}. The NIST AI RMF also refers to the need for implementing post-deployment monitoring plans that capture and evaluate “input from users and other relevant AI actors, appeal and override, decommissioning, incident response, recovery, and change management” \cite{nist2023artificial}. Moreover, the FTC’s Advice on AI Claims suggests that the FTC’s investigations can also act as a form of post-deployment monitoring and intervention \cite{atleson2023keep}. In China, the TC260 Basic Safety Requirements for Generative Artificial Intelligence Services refer to the need for monitoring and evaluation that discover safety issues in the process of service provision and also generally stipulate a range of pre-deployment evaluations that service providers must conduct \cite{sac2024basic}. 

Both sides also agree on the need for standardization of the evaluations that are run. The Biden AI EO emphasizes the need for ``robust, reliable, repeatable and standardized evaluations of AI" \cite{eo2023executive}. The Standardization Administration of China calls for standards to be developed for ``technical requirements and evaluation methods for artificial intelligence" \cite{miit2024guidelines}.

However, the content of evaluations appears oto be different, though with increasing convergence. The Biden AI EO makes clear the need for evaluations of concerning capabilities such as CBRN threats and cyber capabilities \cite{eo2023executive}. As mentioned previously, there is increasing awareness of such risks from AI in China (e.g., via mention of this in the TC260 GenAI Safety Requirements), but the closest evaluations run with concerning or dangerous capabilities in mind are a safety benchmark by the CAICT, which includes hazardous chemicals as a risk under ‘content security’ (renamed to bottom lines/red lines in a later update) \cite{ding2024chinai}. This emphasizes the risk posed by the content generated by AI, rather than a border assessment of whether the AI makes it easier to build, for example bioweapons, than would be the case without AI. That said, Chinese companies are aware of the model evaluations being conducted in the U.S., as these are referenced in the Tencent AI Governance Whitepaper \cite{tencent2024large}. Baichuan includes alignment and capabilities evaluations, though not dangerous capabilities, in their technical papers \cite{baichuan2023baichuan, concordia2024china}. 

Interestingly, the U.S. documents also make greater mention of the use of AI for evaluations and the defensive advantage it confers compared to Chinese documents. The CBRN threats assessment that the Biden AI EO calls on the Department of Homeland Security to conduct explicitly calls for consideration of the defensive advantages AI confers \cite{eo2023executive}. Finally, the U.S. documents also appear interested in the continued monitoring of algorithmic decision-making, monitoring it especially for bias. This is brought up by the Equal Employment Opportunity Committee and the Biden AI EO \cite{eo2023executive}.

\subsubsection{Adversarial testing:} There is some similarity in the adversarial testing required and conducted in both the U.S. and China. In the U.S., however, the focus is in large part on dual-use foundation models, whilst in China, the requirements apply generally to all kinds of AI systems. In both the U.S. and China there is a government requirement for adversarial testing and red-teaming to take place. The Biden AI EO calls for relevant agencies to establish guidelines to enable developers of AI, especially dual-use foundation models, to conduct AI-red teaming campaigns \cite{eo2023executive}. In China, the onus appears to be on AI service providers, as a machine learning security assessment standard (GB/T 42888-2023) calls on service providers to conduct adversarial testing across many scenarios ranging from situations where attackers have only access to model inputs and outputs (black-box) to where they have full control of the model internals (white-box) \cite{sac2023information}. 

There is voluntary industry self-governance corporate compliance with respect to adversarial testing and red-teaming in both jurisdictions as well. The GPT-4, Claude 3, and Gemini technical reports refer to red-teaming and adversarial testing for vulnerabilities, social harms, and dangerous capabilities \cite{google2023gemini, anthropic2024claude, openai2023gpt}. In China, a Tencent whitepaper covering safety and security mentions unique risks of AI systems, which they manage in part through an automated prompt security platform that simulates a range of adversarial attacks \cite{tencent2024large}. They also acknowledge the need for red teams to include a diversity of attack methods and risk scenarios in their evaluations of models. 

\subsubsection{External auditing:} There appears to be some overlap here although 3rd party involvement is seen as favorable or desired in the U.S., while it is seen as just another option service providers can use in China if they lack internal capacity. In the U.S., through the White House voluntary commitments, companies commit to external security testing, independent evaluations, and facilitating third-party discovery and disclosure of security vulnerabilities \cite{whitehouse2023fact}. This appears to be in part driven by a desire to have independent auditing and evaluation capacity, and in part because issues that persist after deployment would more likely be discovered by third parties. In China, on the other hand, an Alibaba whitepaper, which states that a “multidimensional security evaluation” should be carried out before use, and that if an AI service provider does not have the capacity to do this themselves, they can use “a neutral third party organization” \cite{alibaba2023white}. In addition, a draft law written by a group of scholars from Northwest University for Politics and Law says that “developers and providers of critical artificial intelligence should conduct AI security risk assessments on critical AI at least once a year on their own or entrust third-party agencies to conduct timely rectification of discovered security issues and report to the artificial intelligence authorities” \cite{cls_ai_law}.

\subsubsection{Licensing or registration:} There is moderate overlap in this category as both jurisdictions are increasingly focused on tiered oversight for more capable AI models implemented through some type of licensing or registration system. The U.S. has notification and reporting requirements in the Biden AI EO (similar in nature to what we define as registration) extending to models that are trained above a certain threshold (1026 FLOPs for general models, 1023 FLOPs for models trained on biological sequence data) \cite{eo2023executive}. The Biden AI EO also calls for new reporting requirements on cloud compute U.S.ge by foreign actors training AI through American Infrastructure as a Service (IaaS) Providers. A separate executive order calls for creation of regulation wherein U.S. persons will need to provide notification of information relative to specific transactions that might be related to investments in national security technologies \cite{EO14105}. 

Currently in China there is also a registration system, although of a different kind, wherein service providers with social mobilization properties used in recommender systems, deep synthesis, and generative AI systems, must file their algorithms, alongside security assessments and other documents with the CAC \cite{cac2021provisions}. AI law proposals from legal scholars and industry in China call for potential registration or licensing oversight regimes, depending on the capability or compute threshold of models, as well as their potential national security or economic impact \cite{cass2024model, cls_ai_law}.

\section{Recommendations}
The consolidated picture that emerges from our analysis is a promising one, where there are several areas of strong and moderate overlap in risk perception and governance approaches across a range of governmental and corporate documents from the U.S. and China. In this section, we synthesize the results of our analysis into a more holistic set of recommendations, including suggestions for actors who may be best placed to carry forward these dialogues. 

\textbf{First, we recommend that U.S. and Chinese governments strengthen existing intergovernmental dialogue, covering issues related to national security, such as evaluating models for relevant capabilities and preventing proliferation to non-state actors.} Both countries emphasize the importance of testing and evaluating AI systems, and are concerned about the national security-relevant capabilities that AI systems may possess. Currently, while the importance of testing and evaluation is shared between both countries, the content of these evaluations is different, with Chinese evaluations being less focused on national security-relevant capabilities of AI models, and more focused (though not exclusively) on control of politically sensitive content. However, there is evidence from Track II processes which suggests that the two governments could build initial consensus on which domains must be tested for and what commonly agreed upon critical thresholds, such as red lines, might be. A subgroup of this dialogue could focus on the risk of advanced AI proliferation to non-state actors, and aim to build consensus on how to limit such proliferation. Both U.S. and Chinese sources stress the importance of preventing model weight theft and a common set of actors that both countries would be concerned about are non-state actors with significant cyber capabilities. 

\textbf{Next, we recommend that a series of technical standards-setting discussions be conducted, either through existing standards bodies (e.g., ISO) or new fora.} There is serious overlap on the need for reliability, robustness, and adversarial testing in both the U.S. and China. The scope of these discussions should be restricted to issues unrelated to national security considerations of AI systems, instead focusing largely on commercial product safety. These topics may prove especially salient as AI companies from both the U.S. and China increasingly sell products internationally.

\textbf{We also recommend that existing industry coordination approaches such as the Coalition for Content Provenance and Authenticity (C2PA) consider either involving Chinese companies in existing dialogues or setting up distinct tracks of dialogue with Chinese actors.} There is common concern about the need for better information traceability and watermarking across both U.S. and Chinese governments and companies. C2PA is an existing industry association that already brings together leading companies in the U.S. to tackle this issue, and could consider working with Chinese companies to build global norms and standards.

\textbf{Finally, we recommend that Track II dialogues on emerging or novel approaches to safety and governance where there may be some existing common ground.} Dialogues focused on governance, such as the Yale Law School and Chinese Academy of Social Science Law Centre’s Track II dialogue \footnote{https://law.yale.edu/yls-today/news/paul-tsai-china-center-promotes-U.S.-china-communication-amid-bilateral-strains-0}, could include discussion on innovative governance methods for AI as part of their existing dialogues. Such dialogues could include more detailed discussion on items such as compute thresholds and model registration systems, exchanging best practices and lessons from implementation. Our analysis suggests that there is at least some common ground on the use of AI for AI safety, such as by aiding in or automating evaluations of new models. Track 2 dialogues focused on convening scientists, such as IDAIS, could leverage this common ground by initiating dialogue on how AI systems can be used to improve AI safety. 

\section{Discussion}
This paper contributes to ongoing efforts to address one of the most challenging and critical international governance issues of our time: fostering meaningful cooperation between the U.S. and China on managing the risks associated with the proliferation of advanced AI systems \cite{bengio2024managing}. While intense competition between these two countries is likely to persist, in this work, we show through a comprehensive review of over 40 primary AI policy and corporate governance documents from both nations that such fundamental disagreements need not preclude cooperation on all fronts.

While our analysis provides a foundational framework for U.S.– China cooperation on AI governance, several limitations warrant acknowledgment. The rapidly evolving nature of AI policy presents a key challenge, with analyzed documents being regularly updated. For instance, China's TC260 Committee's generative AI safety standard has undergone significant revisions during our study period. Additionally, our focus on document analysis, while enabling systematic comparison, may not fully capture implementation nuances or unpublished contextual information that influence policy formation. Our scope's restriction to high-level domestic policies and AI-specific corporate documents may also overlook relevant convergences in sub-national policies or adjacent regulatory domains, such as China's Personal Information Protection Law. Finally, our identification of common ground does not necessarily indicate areas where dialogue would be most valuable, nor does it constitute our endorsement of specific governance approaches. To address these limitations, we encourage future work to document case studies of dialogue implementation successes and challenges, as well as conduct expert interviews to further understand cooperative efforts on U.S.-China AI relations.

It is also worth re-iterating that our analysis on the U.S. end heavily features the Biden AI EO, which has now been rescinded. Looking at the transition between the first Trump administration and the Biden administration, we observed continuity in AI policy approaches, particularly in areas like global standards setting and national security priorities. However, exact implementation details and enforcement priorities will likely evolve in ways that may be difficult to fully predict based only on historical patterns.

Despite these limitations, our analysis offers a substantive foundation for understanding high-level governmental and corporate priorities in AI governance in the U.S. and China. Given limited existing work in this area, we believe our work will be a critical resource for researchers, as well as practitioners, who are looking for grounded recommendations on topics to include in upcoming dialogues and future research collaborations.

\begin{acks}
Please note that mentioning individuals and organizations here does not imply their endorsement of this paper's content. We are grateful for the input and feedback on earlier drafts of this work provided by the following individuals: Boxi Wu, Conor McGurk, Jason Zhou, Karson Elmgren, Kayla Blomquist, Oliver Guest, and Scott Singer. We are especially thankful to Zachary Arnold from the Center for Security and Emerging Technology for sharing the taxonomy from the AGORA project with us and allowing us to adapt it for this project. 
\end{acks}
\appendix
\section{List of Acronyms}

\subsection*{General}
\begin{description}
    \item[UNCLOS] UN Convention on the Law of the Sea
    \item[EEZ] Exclusive Economic Zone
    \item[IDAIS] International Dialogues on AI Safety
    \item[UNGA] The United Nations General Assembly
    \item[IP] Intellectual Property
    \item[CBRN] Chemical, Biological, Radiological or Nuclear
    \item[FLOPS] Floating Point Operations Per Second
    \item[IaaS] Infrastructure as a Service
    \item[C2PA] Coalition for Content Provenance and Authenticity
    \item[PETs] Privacy-Enhancing Technologies
\end{description}

\subsection*{U.S.}
\begin{description}
    \item[FTC] Federal Trade Commission
    \item[CSET] Center for Security and Emerging Technology
    \item[NSA] National Security Agency
    \item[NIST] National Institute of Standards and Technology
    \item[RMF] NIST's AI Risk Management Framework
    \item[DHS] Department of Homeland Security
    \item[DOE] Department of Energy
    \item[DHHS] Department of Health and Human Services
    \item[DPA] Defense Production Act
    \item[CISA] Cybersecurity and Infrastructure Security Agency
    \item[EO] Executive Order
    \item[AI EO] Biden administration's 2023 ``Executive Order on the Safe, Secure, and Trustworthy Development and Use of Artificial Intelligence''
\end{description}

\subsection*{China}
\begin{description}
    \item[SAC] Standardization Administration of China
    \item[CCP] Chinese Communist Party
    \item[TC260] National Information Security Standardization Technical Committee
    \item[CAICT] China Academy of Information and Communications Technology
    \item[MIIT] The Ministry of Industry and Information Technology
    \item[CAC] The Cyberspace Administration of China
    \item[CASS] Chinese Academy of Social Science
    \item[CUPL] China University of Political Science and Law
    \item[CISS] CISS Center for International Security and Strategy of Tsinghua University
\end{description}
\section{Taxonomies} 
\begin{table*}[!htbp]
\caption{Taxonomy of risks analysed in AI documents from the U.S. and China. The definitions are adapted from the AGORA database with one additional category: dangerous capabilities.}
\label{tab:risks}
\begin{tabular}{lp{0.6\textwidth}}
\toprule
\textbf{Risk} & \textbf{Definition} \\
\midrule
Limited user transparency & Risks related to scenarios where individuals or entities do not have access to information about inputs into particular decisions and/or critical details about the AI system they are interacting with \\
\midrule
Poor reliability & Risks related to the inability of an AI system to perform as required, under the conditions of expected use and over a given period of time, including the entire lifetime of the system \\
\midrule
Lack of robustness & Risks related to an AI system's inability to function as intended under unexpected or unusual circumstances, such as when encountering adversarial inputs or data outside the training distribution \\
\midrule
Bias and discrimination & Risks related to undesirable biases in the outputs of AI systems, including biases according to commonly protected classes such as race or gender \\
\midrule
Lack of interpretability and explainability & Risks stemming from opacity of mechanisms underlying an AI system's operation and/or the meaning of the systems' output in the context of use \\
\midrule
Dangerous capabilities & Risks related to AI system capabilities that could pose critical, large-scale, and national security-relevant threats, including CBRN, cyber, persuasion, autonomy and self-replication \\
\midrule
Weak cybersecurity & Risks related to the security of digital systems associated with AI, including systems involved in development and training, systems housing related intellectual property, and computing infrastructure for deployed models \\
\midrule
Lack of privacy & Risks related to the unauthorized use, disclosure or sharing of personally identifying information in an AI system's inputs or outputs \\
\bottomrule
\end{tabular}
\end{table*}

\begin{table*}[h]
\caption{Taxonomy of governance approaches analyzed in AI documents from the U.S. and China. The definitions are adapted from the AGORA database with one additional category: use of AI for pro-safety purposes.}
\label{tab:governance}
\begin{tabular}{lp{0.6\textwidth}}
\toprule
\textbf{Governance approach} & \textbf{Definition} \\
\midrule
Use of AI for pro-safety purposes & Using AI systems to improve AI safety and general societal resilience (e.g., using AIs for detection of anomalies) \\
\midrule
Convening & Facilitating, requiring, setting conditions on, or otherwise addressing the convening of different stakeholders to tackle governance challenges - for example, to share feedback or to participate in its development or deployment \\
\midrule
Governance development & Creating, supporting, or requiring the development of public or private governance mechanisms or institutions related to the development or deployment of AI systems \\
\midrule
Technical mitigations & Developing technical solutions to manage risks - for example, watermarking and content labeling \\
\midrule
Evaluations & Requiring, or encouraging, the systematic evaluation of AI systems \\
\midrule
Adversarial testing & Evaluation in which the evaluator takes an adversarial approach, seeking to subvert or otherwise produce undesirable results from the system or process being tested by any means available. Sometimes called "red teaming" \\
\midrule
External auditing & Evaluation by a disinterested counterparty or third party, such as a customer or a professional auditing firm \\
\midrule
Licensing or registration & Requiring, incentivizing, or otherwise encouraging actors involved in AI-related activities, such as AI developers, vendors, users, or researchers, to either receive sanction from a regulator for their activities (licensing) or to notify a regulator of their activity pursuant to a formal process (registration) \\
\midrule
Risk assessments & Refers to the identification, assessment and in some cases quantification of potential sources of and pathways to harm caused by AI systems \\
\midrule
Disclosure & Requiring or encouraging, the disclosure of information about AI systems by their users, developers, vendors, or other stakeholders to third parties, including but not limited to the general public \\
\midrule
Input controls & Restricting or placing conditions on the sale, distribution, or use of technical inputs to AI systems, specifically data or computational resources \\
\midrule
Liability & Holding specific parties accountable through civil or criminal penalties for unlawful actions \\
\midrule
Pilot and testbeds & Creating, facilitating, setting conditions on, or otherwise addressing the development and operation of government-supported or government-conducted pilot programs or test environments related to artificial intelligence \\
\bottomrule
\end{tabular}
\end{table*}
\bibliographystyle{acm}
\bibliography{sample-base}
\end{document}